\documentclass[
 reprint,
superscriptaddress,
 amsmath,amssymb,
 aps,
 prx,
]{revtex4-2}
\usepackage{changes}
\usepackage{soul}

\usepackage{changes}
\usepackage{soul}
\usepackage[utf8]{inputenc}
\usepackage[english]{babel}
\usepackage{amsmath,amsfonts,amssymb}
\usepackage[T1]{fontenc}
\usepackage{graphicx}
\usepackage[colorlinks=true,bookmarks=false,citecolor=blue,urlcolor=blue]{hyperref}

\usepackage{upgreek}

\usepackage{bm}

\graphicspath{{./img/}}

\newcommand{\sini}{$\mathrm{Si_3N_4}$}

\renewcommand*{\thefootnote}{\alph{footnote}}

\usepackage{enumitem}
\setlist{nolistsep}

\newcommand\s[1]{\mathrm{#1}}

\usepackage{letltxmacro}
\makeatletter
\let\oldr@@t\r@@t
\def\r@@t#1#2{%
	\setbox0=\hbox{$\oldr@@t#1{#2\,}$}\dimen0=\ht0
	\advance\dimen0-0.2\ht0
	\setbox2=\hbox{\vrule height\ht0 depth -\dimen0}%
	{\box0\lower0.4pt\box2}}
\LetLtxMacro{\oldsqrt}{\sqrt}
\renewcommand*{\sqrt}[2][\ ]{\oldsqrt[#1]{#2}}
\makeatother

\begin{document}
\title{Observation of stimulated Brillouin scattering in silicon nitride integrated waveguides}

\author{Flavien Gyger}
\thanks{These authors contributed equally to this work}
\affiliation{Group for Fibre Optics, Swiss Federal Institute of Technology Lausanne (EPFL), CH-1015 Lausanne, Switzerland}

\author{Junqiu Liu}
\thanks{These authors contributed equally to this work}
\affiliation{Laboratory of Photonics and Quantum Measurements, Swiss Federal Institute of Technology Lausanne (EPFL), CH-1015 Lausanne, Switzerland}

\author{Fan Yang}
\thanks{These authors contributed equally to this work}
\affiliation{Group for Fibre Optics, Swiss Federal Institute of Technology Lausanne (EPFL), CH-1015 Lausanne, Switzerland}

\author{Jijun He}
\affiliation{Laboratory of Photonics and Quantum Measurements, Swiss Federal Institute of Technology Lausanne (EPFL), CH-1015 Lausanne, Switzerland}

\author{Arslan S. Raja}
\affiliation{Laboratory of Photonics and Quantum Measurements, Swiss Federal Institute of Technology Lausanne (EPFL), CH-1015 Lausanne, Switzerland}

\author{Rui Ning Wang}
\affiliation{Laboratory of Photonics and Quantum Measurements, Swiss Federal Institute of Technology Lausanne (EPFL), CH-1015 Lausanne, Switzerland}

\author{Sunil A. Bhave}
\affiliation{OxideMEMS Lab, Purdue University, 47907 West Lafayette, IN, USA}

\author{Tobias J. Kippenberg}
\email{tobias.kippenberg@epfl.ch}
\affiliation{Laboratory of Photonics and Quantum Measurements, Swiss Federal Institute of Technology Lausanne (EPFL), CH-1015 Lausanne, Switzerland}

\author{Luc Th{\'e}venaz}
\email{luc.thevenaz@epfl.ch}
\affiliation{Group for Fibre Optics, Swiss Federal Institute of Technology Lausanne (EPFL), CH-1015 Lausanne, Switzerland}

\begin{abstract}
Silicon nitride (Si$_3$N$_4$) has emerged as a promising material for integrated nonlinear photonics and has been used for broadband soliton microcombs and low-pulse-energy supercontinuum generation. Therefore understanding all nonlinear optical properties of Si$_3$N$_4$ is important. 
So far, only stimulated Brillouin scattering (SBS) has not been reported. Here we observe, for the first time, backward SBS in fully cladded Si$_3$N$_4$ waveguides. The Brillouin gain spectrum exhibits an unusual multi-peak structure resulting from hybridization with high-overtone bulk acoustic resonances (HBARs) of the silica cladding. 
The reported intrinsic \sini\ Brillouin gain at 25 GHz is estimated as $7\times10^{-13}$ m/W.
Moreover, the magnitude of the \sini\ photoelastic constant is estimated as $|p_{12}|=0.047\pm 0.004$. Since SBS imposes an optical power limitation for waveguides, our results explain the capability of Si$_3$N$_4$ to handle high optical power, central for integrated nonlinear photonics.
\end{abstract}

\maketitle

\textit{Introduction - }Integrated photonics has significantly advanced over the past decades. Today, integrated photonics is used to build on-chip lasers which can be found in data centers, and passive optical elements such as filters and arrayed waveguide gratings for optical signal processing. Although silicon and indium phosphide are the most mature platforms, 
there has been growing interest and advances in silicon nitride (\sini).
Amorphous \sini\ \cite{moss_new_2013} shows exceptional performance in terms of low linear optical losses below 1 dB/m and absence of two-photon absorption, and has been widely used for passive elements such as delay lines and multi-mode interferometers \cite{Blumenthal:18}. In addition, the high Kerr nonlinearity and flexibility to engineer the anomalous group velocity dispersion (GVD) via geometry variation \cite{Foster:06} have made \sini\ an ideal platform for integrated nonlinear photonics. Moreover, \sini\ is suitable for applications in space \cite{Brasch:14}. 
Although already considered in the 1980's for its promise in integrated photonics \cite{Henry:87}, only 
recent advancements in nanofabrication for film growth and patterning have overcome the highly tensile film stress of stoichiometric \sini\ \cite{Luke:13, Pfeiffer:18a}. These developments allow high-yield fabrication of \sini\ waveguides with tight optical confinement and anomalous GVD, as required for parametric frequency conversion via Kerr nonlinearity \cite{Kippenberg:04}.
Such integrated \sini\ waveguides are presently a leading platform for dissipative-Kerr-soliton-based frequency comb ("soliton microcomb") generation \cite{kippenberg_dissipative_2018}. Integrated Si$_3$N$_4$-based soliton microcomb can now operate with ultralow electrical driving power \cite{stern_battery-operated_2018, Raja:19} and repetition rate extending down into the microwave domain (e.g. X- and K-band) \cite{liu_nanophotonic_2019}, and have been used for system-level demonstrations \cite{kippenberg_dissipative_2018}. 
In addition, \sini\ waveguides enables coherent and low-pulse-energy supercontinuum generation in near-infrared \cite{gaeta_photonic-chip-based_2019}, as well as in mid-infrared for dual-comb spectroscopy \cite{Guo:19}.
Other nonlinear phenomena in \sini\ waveguides such as second- and third-harmonic generation \cite{Levy:11, Xue:17}, as well as stimulated Raman scattering (SRS) \cite{Karpov:16} have been studied. However to date, there is no report on backward Brillouin scattering in \sini. Understanding the acousto-optic interaction in \sini\ waveguides is key to characterize Brillouin gain, and equally important,
to model the noise properties resulting from thermally-excited guided acoustic waves.

Stimulated Brillouin scattering (SBS) is a nonlinear process mediated by acousto-optic interaction inside a medium \cite{eggleton_brillouin_2019,safavi-naeini_controlling_2019,wiederhecker_brillouin_2019}. It has been observed in various platforms, including silica fibers \cite{ponikvar_stabilized_1981,kang_tightly_2009,pang_stable_2015,beugnot_brillouin_2014,florez_brillouin_2016}, whispering-gallery-mode resonators \cite{grudinin_brillouin_2009,tomes_photonic_2009,lee_chemically_2012,yang_bridging_2018}, and integrated waveguides based on chalcogenide \cite{pant_-chip_2011,morrison_compact_2017}, silicon \cite{shin_tailorable_2013,van_laer_interaction_2015} and aluminum nitride \cite{liu2019electromechanical}. SBS has led to several applications, such as slow and fast light \cite{song_observation_2005,okawachi_tunable_2005}, microwave photonic filters \cite{shin_tailorable_2013,marpaung_low-power_2015}, microwave synthesis \cite{li_electro-optical_2014}, highly coherent laser sources \cite{loh_microrod-resonator_2016,gundavarapu_sub-hertz_2019}, gyroscopes \cite{li_microresonator_2017,gundavarapu_sub-hertz_2019}, isolators \cite{kang_reconfigurable_2011}, mode-locked lasers \cite{pang_stable_2015,pang_all-optical_2016} and sensors \cite{antman_optomechanical_2016}.
At the same time, Brillouin scattering poses a power limitation in waveguides and fibers \cite{Smith:72}, and in addition induces noises via thermal excitation of guided acoustic waves.

Recently, integrated Si$_3$N$_4$ waveguides have been used to demonstrate a Brillouin laser \cite{gundavarapu_sub-hertz_2019}. However in that work, Si$_3$N$_4$ was used solely to guide the light, and the SBS interaction occurred within the silica (SiO$_2$) cladding, as confirmed by the Brillouin frequency shift of 10.9 GHz. A second work proposed the use of Si$_3$N$_4$ membrane including a waveguide and a phononic crystal to explore forward SBS \cite{dehghannasiri_raman-like_2017,dehghannasiri_observation_2017}. A third work demonstrated a large light-sound interaction using forward SBS in integrated silicon waveguides \cite{shin_tailorable_2013}, while a \sini\ membrane was used to guide the transverse phonons. However, to date, no work has shown backward SBS in \sini\ material. Moreover, to the best of our knowledge, there is no reference to the \sini\ photoelastic tensor (e.g. $p_{11}$ and $p_{12}$) in the literature. 

\begin{figure}
		\includegraphics[width = 8.5cm]{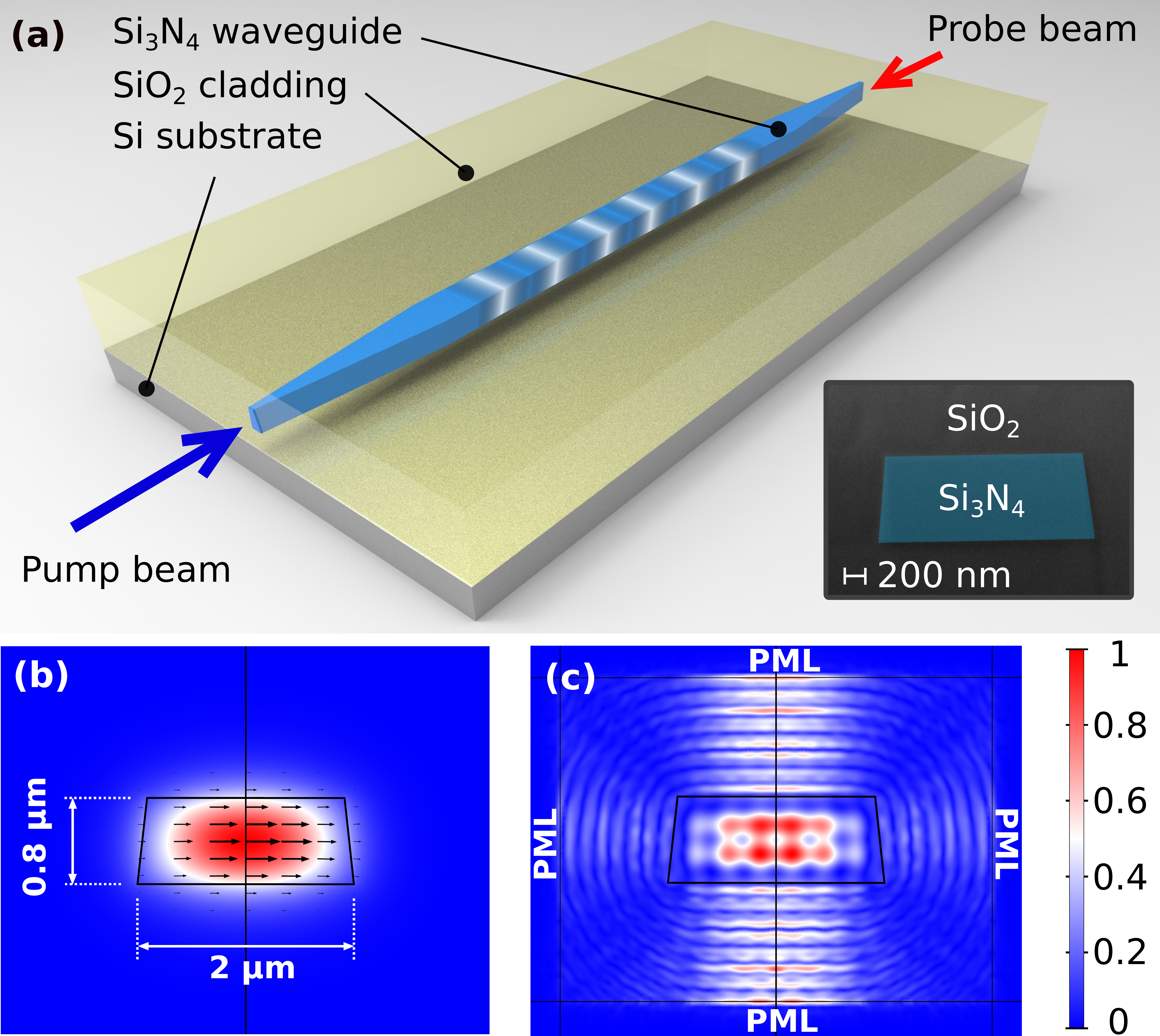}
		\caption{\textbf{On-chip SBS in a $\bm{\mathrm{Si_3N_4}}$ waveguide.}\\\textbf{(a)} Artist's view of SBS in a \sini\ waveguide. Note that the length of the inverse tapers is exaggerated for illustrative purpose. The actual taper length is 300 $\upmu$m, while the full length of the \sini\ waveguide is 5 mm. White and blue shading represents the material density variation generated by SBS when the pump and probe beams overlay. Inset: false-color SEM of the waveguide cross-section of $2\times0.8$ $\upmu$m$^2$. \textbf{(b)} Normalized electric field distribution for the optical TE$-$like mode. The arrows represent the direction and strength of the electric field in the cross-section plane. \textbf{(c)} Normalized displacement field norm for the fundamental acoustic mode. Visible waves exiting the waveguide indicate a rather high phonon leakage rate. A PML is surrounding the entire domain and absorbs these waves. The vertical black line visible in \textbf{(b)} and \textbf{(c)} is indicative of the waveguide symmetry.}
		\label{fig_illustration}
\end{figure}

In this Letter, we characterize, for the first time, the \sini\ backward SBS gain spectrum in a 5-mm-long waveguide buried in SiO$_2$. Figure \ref{fig_illustration}(a) shows an artist's view of SBS in a \sini\ waveguide buried in SiO$_2$. Two optical fields, pump and probe, spatially overlap and are phase-matched with an acoustic mode, which induce acoustic oscillations through electrostriction and subsequent generation of a moving grating. Reciprocally, acoustic waves scatter light via photoelasticity. The phase-matching condition is given by:
\begin{equation}
\nu_\s{B} = \frac{2n_{\s{eff}}v_a}{\lambda},
\label{eq_phase_matching}
\end{equation}
where $n_{\s{eff}}$ is the effective refractive index of the optical mode, $\lambda$ is the probe wavelength, $v_a$ is the acoustic mode velocity and $\nu_\s{B}$ is the Brillouin frequency shift. In the present case, our waveguide cross-section is a $2\times0.8$ $\upmu$m$^2$ trapezoid, exhibiting a smaller width at the top (see Supplementary Material), as shown in the false-color scanning electron microscopy (SEM) image in Fig. \ref{fig_illustration}(a) inset. The refractive indices of \sini\ and SiO$_2$ at 1550 nm are $n_{\s{Si_3N_4}}= 2.00$ and $n_{\s{SiO_2}}= 1.45$, respectively.

The waveguide fundamental quasi-transverse electric (TE$-$like) mode is adiabatically excited by inverse nanotapers \cite{Almeida:03, Liu:18} placed at chip facets. The simulated optical mode profile in the waveguide using finite-element method (FEM) is shown in Fig. \ref{fig_illustration}(b). The computed effective refractive index and mode area are $n_\s{eff} = 1.85$ and $A_\s{eff}=$ 1.2 $\upmu$m$^2$, respectively. 
A solid mechanics simulation of the fundamental acoustic mode displacement field norm is shown in Fig. \ref{fig_illustration}(c). The details about the simulations are provided in the Supplementary Material. Due to the weak confinement caused by the lower acoustic velocity in SiO$_2$ and by the small waveguide dimension, acoustic waves propagate away from the waveguide and are absorbed by the surrounding perfectly matched layer (PML) \cite{eggleton_brillouin_2019,poulton_acoustic_2013}. Moreover, the mode shape in the waveguide substantially deviates from the usual bell shape. This is caused by the small waveguide dimensions inducing hybrid acoustic modes \cite{dainese2006stimulated}. Due to the negligible optical evanescent field in SiO$_2$ cladding, SBS happens mainly in \sini. Indeed, our simulations show that the SiO$_2$ contributes to $0.2\%$ of the total Brillouin gain of the aforementioned acoustic mode.

\begin{figure*}[t!]
	\includegraphics[width = 17cm]{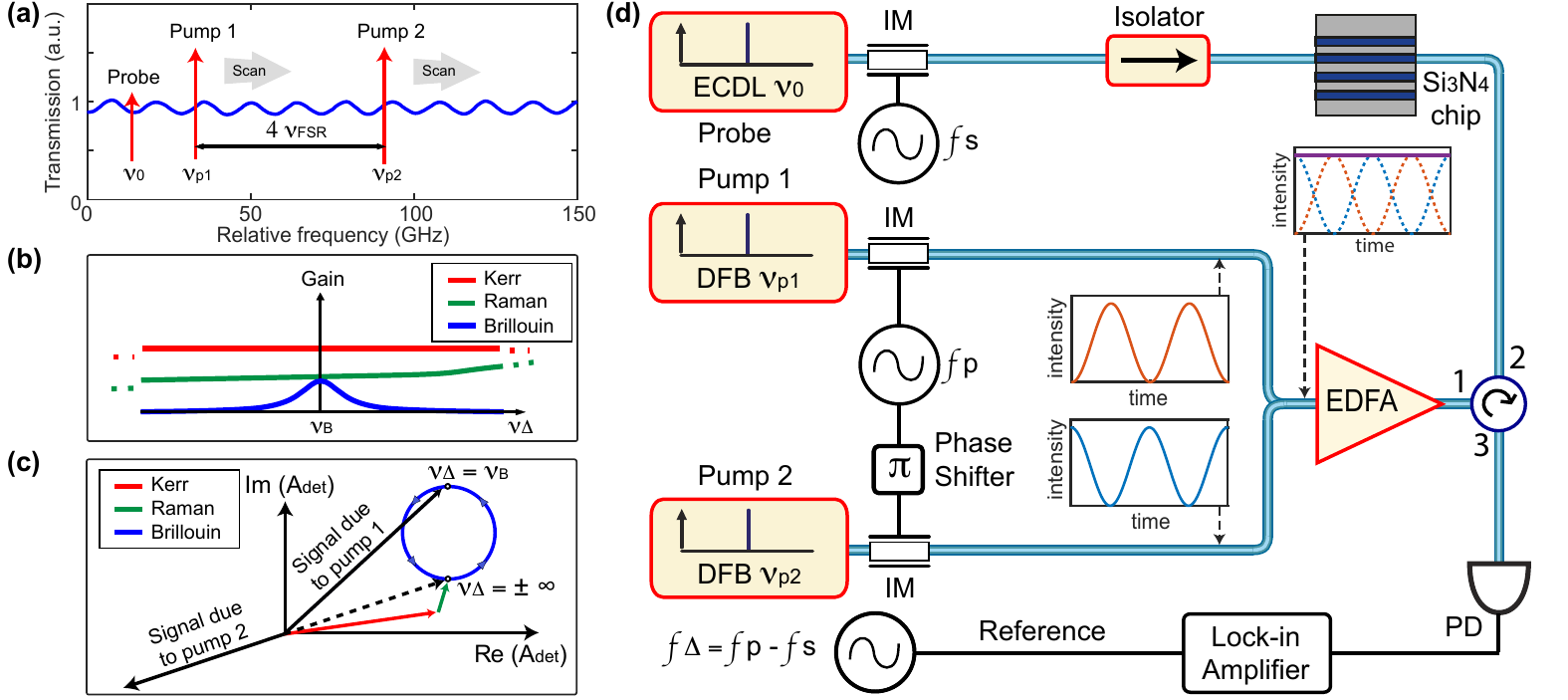}
	\caption{\textbf{High sensitivity SBS gain measurement scheme.} \textbf{(a)} Measured waveguide transmission spectrum in function of the relative frequency starting at 193.55 THz. The frequency difference of the two pumps, $\nu_\s{p2}-\nu_\s{p1}$, is a multiple of the chip FP cavity's FSR ($4\cdot\nu_\s{FSR}$). The two pumps are scanned together while the probe frequency $\nu_0$ is fixed. \textbf{(b)} Illustration of the gains due to Kerr effect (phase-to-intensity conversion) and SRS nearby the Brillouin gain, showing a rapid Brillouin gain change while Kerr and SRS contributions are nearly flat. \textbf{(c)} Illustration in the complex-plane of different contributions of the measured probe signal amplitude $A_\s{det}$. The signal generated by pump 1 is a sum of Kerr effect, SRS and SBS, while the signal generated by pump 2 only contains Kerr effect and SRS. Moreover, these contributions are $\pi$-phase-shifted, leading to a cancellation of the background signal. Note that the position of pump 1 has been drawn at an arbitrary position nearby the Brillouin peak. Moreover, the phase shifts of Kerr effect, SRS and SBS have been drawn arbitrarily, which depend on the precise experiment configuration. \textbf{(d)} Simplified TIM experimental setup. ECDL, external-cavity diode laser; DFB laser, distributed feedback laser; EDFA, erbium-doped fiber amplifier; IM, intensity modulator; PD, photodetector.}
	\label{fig_TIMsetup}
\end{figure*}

\textit{Experiments and results - }
Residual reflections between the two chip facets make the waveguide a low-finesse Fabry-Pérot (FP) cavity. The transmission spectrum of the waveguide is shown in Fig. \ref{fig_TIMsetup}(a), exhibiting a FP cavity free spectral range (FSR) of $\nu_\s{FSR} = 14.375$ GHz, corresponding to the 5-mm waveguide length.
This cavity causes several experimental challenges: (1) The probe transmission is intensity-modulated by low-frequency random environmental fluctuations (e.g. temperature changes due to variations in the coupling ratio of the pump beam), which constantly shifts the FP cavity's transmission spectrum. In our experiments, this process generates a noise of standard deviation $10^5$ times higher than the Brillouin signal.
(2) Intensity modulation of the pump signal modulates the cavity refractive index via Kerr effect. The resulting time-dependent shift of the FP cavity transmission spectrum modulates the probe signal. Hence, pump intensity variations are transferred to the probe signal and generate a noisy background of standard deviation two times larger than the Brillouin peak gain.
(3) In our work, the chip is coupled via two 1-m-long standard single-mode fiber patchcords \cite{Raja:19a} in which the pump and probe counter-propagate and, as a result, SRS occurs along those fibers and generates a noisy background signal. This process contributes to a lesser extent to the system noise ($\approx$ 10\% of the Brillouin peak gain). Note that \sini\ SRS is considered much smaller than for silica \cite{Karpov:16} and can be neglected.
Details about the estimates given above are provided in the Supplementary Material.

To overcome the three aforementioned challenges, we developed a novel technique, \textit{triple intensity modulation} (TIM), able to measure the exact Brillouin gain profile. This technique is now explained in details. The pump and probe beams are intensity-modulated at frequency $f_\s{P}$ and $f_\s{S}$ respectively, and the detection is made at the frequency difference $f_\Delta = f_\s{P}-f_\s{S}$ (in our experiment, $f_\s{S} =$ 20 MHz and $f_\Delta$ = 75 kHz). This implementation exploits the nonlinear nature of SBS leading to sum-difference frequency generation, and has two benefits: (1) It moves the signal detection frequency away from the direct-current (DC) frequency, where the low-frequency environmental noises lie. (2) It efficiently filters the pump reflection out from the probe beam in the radio-frequency (RF) domain. The filtering is achieved by detecting the signal at a distant frequency from the modulated pump frequency via a lock-in amplifier, eliminating the need for high extinction optical filtering. Such a technique is commonly used to filter out stray light in Brillouin microscopy \cite{grubbs_high_1994}.

The aforementioned Kerr effect and Raman scattering issues are resolved by cancelling the temporal variation of the pump intensity. A similar method has been developed to cancel the non-resonant background in coherent anti-Stokes Raman spectroscopy \cite{Marowsky1990}. This is realized by adding a second modulated pump beam (i.e. pump 2), whose intensity and polarization match perfectly those of the first pump (i.e. pump 1). The optical frequency of pump 2 ($\nu_\s{P2}$) is increased compared to that of pump 1 ($\nu_\s{P1}$) by a multiple of the chip FP cavity FSR: $\nu_\s{P2}=\nu_\s{P1}+n\cdot\nu_{\s{FSR}}$, $n\in\mathbb{N}$ (in our experiment, n = 4), as illustrated in Fig. \ref{fig_TIMsetup}(a). In this way, both pumps' intensities remain identical at all times inside the chip, irrespective of environmental noises. In addition, the two pumps are intensity-modulated at the same frequency $f_\s{P}$ but are $\pi$-phase shifted, such that perfect intensity cancellation of the modulation frequency $f_\s{P}$ occurs. Hence, the Kerr effect no longer modulates the probe transmission, due to the constant total pump intensity. Moreover, the SRS gain experienced by the probe is nearly identical for each pump, because the frequency difference of the two pumps, $\nu_\s{P2}-\nu_\s{P1}$, is much smaller than the silica Raman gain bandwidth ($\sim$ 7 THz). 

Since the two pumps are modulated with $\pi$-phase difference, SRS is attenuated at the detection frequency $f_\Delta = f_\s{P}-f_\s{S}$. For SBS, however, the probe only interacts with pump 1. The situation is summarized in Fig. \ref{fig_TIMsetup}(b) in which the gain behaviour of the Kerr effect (phase-to-intensity conversion can be seen as a gain process for the probe beam), SRS and SBS are sketched in function of the pump-probe detuning frequency $\nu_\Delta = \nu_\s{P1}-\nu_{0}$, where $\nu_{0}$ is the probe frequency. It can be seen that Kerr effect is flat over the entire detuning frequency range and the Raman gain of the silica patchcord is nearly flat over the Brillouin gain bandwidth. Thus, these two contributions result in an identical gain for the two pumps. As the pumps have a $\pi$-phase difference, these effects cancel out at the detection frequency $f_\Delta$. A different view of this cancellation is represented in Fig. \ref{fig_TIMsetup}(c) where the different contributions of the probe detection signal $A_\s{det}$, are represented in the complex-plane, for the two pumps. Kerr effect and SRS generate a background signal (dotted arrow) that is cancelled by pump 2. The blue circle represents the path of the SBS Lorentzian amplitude lineshape that is traced when the pump-probe detuning frequency $\nu_{\Delta}$, is scanned. Brillouin gain spectral shape is measured by scanning the two pump frequencies $\nu_\s{P1}$ and $\nu_\s{P2}$, while fixing the probe frequency $\nu_0$. The two pump frequencies are scanned simultaneously such that their frequency difference $\nu_\s{P2}-\nu_\s{P1}$ remains constant during the scanning, as illustrated in Fig. \ref{fig_TIMsetup}(a).

The experimental setup is shown in Fig. \ref{fig_TIMsetup}(d). The two pumps are generated by two distributed feedback (DFB) lasers and the frequency scanning is performed by temperature control of these lasers. In the small gain approximation \cite{boyd_book}, the magnitude of the measured amplitude $|A_{\s{det}}|$ in the presence of SBS in the \sini\ waveguide can be written as \cite{gyger2018high}:
\begin{equation}
|A_{\s{det}}(\nu_\Delta)| \propto \rho_\s{pd} \cdot P_\s{S} \cdot P_\s{P} \cdot g_\s{B} \cdot L \cdot \mathcal{L}(\nu_\Delta),
\end{equation}
where $\rho_\s{pd}$ is the photodetector power-to-voltage conversion factor, $P_\s{S}$ is the time-averaged detected probe power, $P_\s{P}$ is the time-averaged pump power in the waveguide (pump 1 only), $g_\s{B}$ is the Brillouin peak gain in unit of $\s{m^{-1}W^{-1}}$, $L$ is the waveguide length, and $\mathcal{L}(\nu_\Delta)$ is the gain lineshape.

\begin{figure*}[t!]
	\includegraphics[width = 17cm]{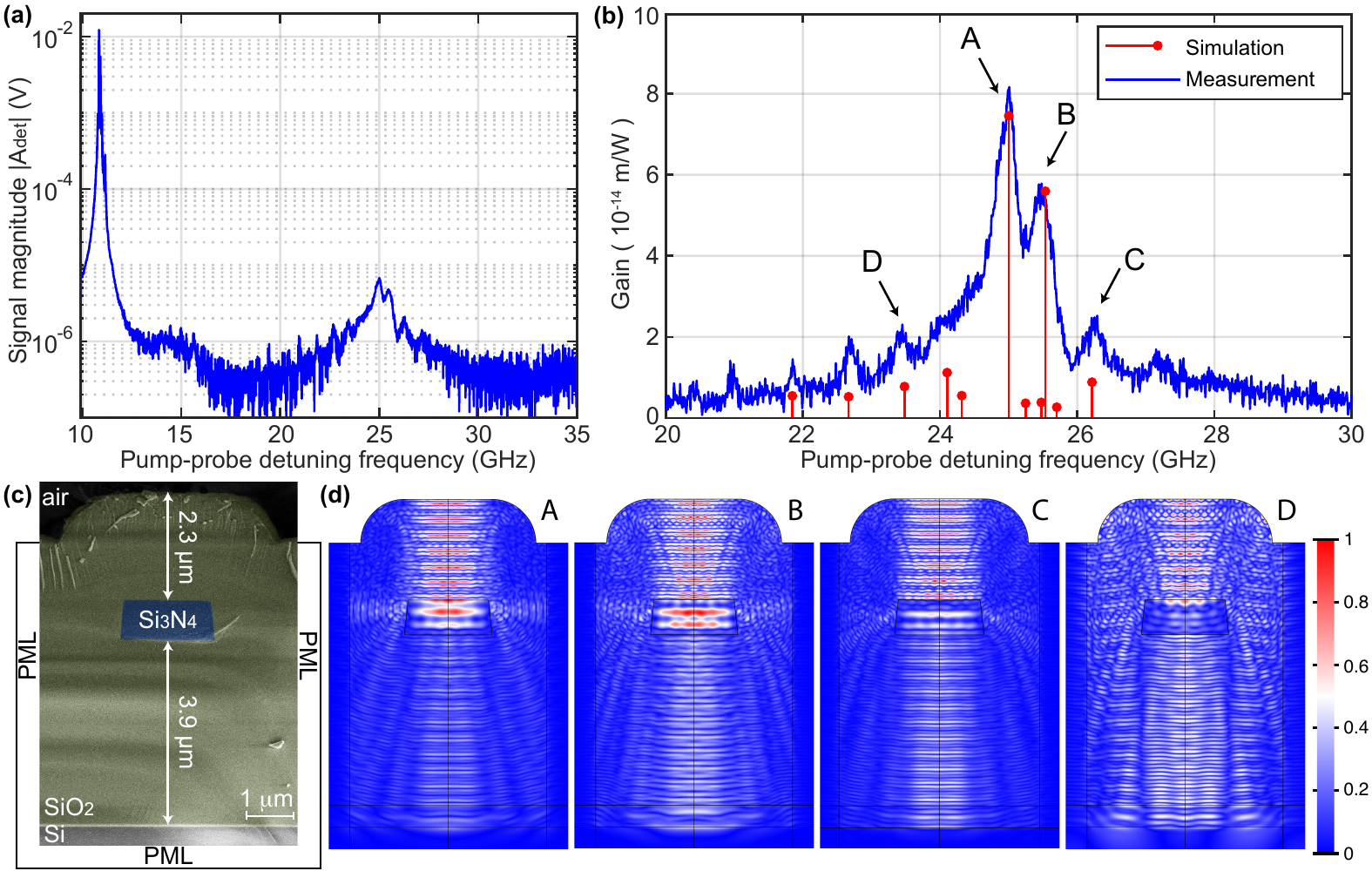}
	\caption{\textbf{Measurement and simulation results of $\bm{\mathrm{Si_3N_4}}$ Brillouin gain spectrum.} \textbf{(a)} Measured signal in function of the pump-probe detuning frequency in logarithmic scale, showing the 1-m-long silica patchcord SBS to the left and \sini\ SBS at 25 GHz, with more than three orders of magnitude difference. 
	\textbf{(b)} Measured \sini\ SBS gain spectrum, in agreement with the simulated eigenmodes (eigenfrequency and gain for each eigenmode). \textbf{(c)} SEM image showing the sample cross-section. The precisely measured geometry parameters are used to build the simulation model.
	\textbf{(d)} Normalized acoustic displacement field norm of four peaks extracted from the simulations. Each peak corresponds to an acoustic supermode resulting from the hybridization with bulk acoustic reflections at the SiO$_2$-air (top) and SiO$_2$-silicon (bottom) boundaries. Note that the top SiO$_2$ dome-shaped boundary is due to fabrication \cite{Xuan:16} (see Supplementary Material).}
	\label{fig_ResultTM}
\end{figure*}

Figure \ref{fig_ResultTM}\textbf{(a)} shows the magnitude of the measured amplitude $|A_\s{det}|$, when the pump-probe detuning frequency is scanned from 10 to 35 GHz. The peak to the left is the silica Brillouin gain of the 1-m-long fiber patchcords connecting to the waveguide, and the signal around 25 GHz is the \sini\ Brillouin gain. Since these two gains are measured jointly, their ratio provides an additional way to estimate the Si$_3$N$_4$ Brillouin gain value. Figure \ref{fig_ResultTM}\textbf{(b)} shows the measured \sini\ gain spectrum along with our simulation results. A main peak reaching $(8\pm1)~\times$ 10$^{-14}$ m/W is found at 25 GHz, accompanied by other smaller peaks. The details about error estimation are provided in the Supplementary Material. Using the phase matching condition Eq. \ref{eq_phase_matching}, the velocity of the main acoustic mode is calculated as 10.5 km/s, which agrees with the literature \cite{Wolff:14}. The full width at half maximum of the main peak $\Delta\nu=390$ MHz, is obtained by a Lorentzian fitting. Figure \ref{fig_ResultTM}\textbf{(d)} shows the FEM simulations performed over a 3D trench of the entire chip cross-section including the SiO$_2$-air top interface and SiO$_2$-silicon-substrate bottom interface. The simulation model is built using the precisely measured parameters from SEM as shown in Fig. \ref{fig_ResultTM}\textbf{(c)}. The details about the simulations are provided in the Supplementary Material. Acoustic eigenmodes, computed from 21 to 27 GHz, show that the multiple peaks in the gain spectrum originate from the hybridization with high-overtone bulk acoustic resonances (HBARs) caused by the reflection of acoustic waves at the SiO$_2$-air top boundary and SiO$_2$-silicon-substrate bottom boundary \cite{Bhave:19}.

\textit{Discussion - } 
In our \sini\ waveguide, the optical mode can be approximated as a transverse mode. Then, the Brillouin gain $\tilde{g}_\s{B}$ in units m/W, can be expressed as:
\begin{equation}
\tilde{g}_\s{B} = \frac{4\pi^2 n^7p_{12}^2\eta}{\lambda^2 c\rho v_a \Gamma},
\label{eq:BrillouinGain}
\end{equation}
where $n$ is the refractive index, $c$ is the speed of light in vacuum, $\rho$ is the material density, $p_{12}$ is the photoelastic constant, $\Gamma$ is the acoustic damping and $\eta$ is the acousto-optic overlap coefficient. We now discuss on the three main parameters: $\Gamma$, $p_{12}$ and $\eta$.

The \textit{acoustic damping} is proportional to the Brillouin linewidth as $\Gamma = \Gamma_M +\Gamma_L = 2\pi\cdot\Delta\nu$, where $\Gamma_M$ represents material damping caused by phonon absorption, and $\Gamma_L$ represents phonon leakage from the waveguide. The \sini\ Brillouin linewidth measured here is 10-times larger than that of silica ($\Gamma_\s{Si_3N_4}=\Gamma_{M, \s{Si_3N_4}}+\Gamma_{L, \s{Si_3N_4}} = 10\cdot\Gamma_{M, \s{SiO_2}}$), leading to a tenfold gain reduction. Note that, material damping generally follows a square dependence on the Brillouin frequency shift ($\Gamma_M \propto \nu_B^2$) \cite{auld1973acoustic}. Compared to 11 GHz silica Brillouin shift, a fivefold increase in material damping is expected in \sini: $\Gamma_{M, \s{Si_3N_4}} = 5\cdot \Gamma_{M, \s{SiO_2}}$. Therefore the remaining damping is assumed to be due to phonon leakage: $\Gamma_{L, \s{Si_3N_4}} = 5\cdot \Gamma_{M, \s{SiO_2}}$.

The \textit{photoelastic constant} quantifies the amount of stress induced in the material by an electric field due to electrostriction. By fitting the peak heights of the FEM simulation results to the measured gain spectrum, the estimated photoelastic constant magnitude $|p_{12}|$, is obtained:
\begin{equation}
|p_{12}|= 0.047\pm 0.004.
\end{equation}
The details about error estimation are provided in the Supplementary Material. This value corresponds to a 5.7-times reduction with respect to that of silica, leading to a 33-times gain decrease as $p_{12, \s{Si_3N_4}}^2= p_{12, \s{SiO_2}}^2/33$.

The \textit{acousto-optic overlap coefficient} is the coupling strength between the optical mode and the acoustic mode in relation to their spatial distributions. For example, almost perfect overlap of optical and acoustic modes is achieved in optical fibers, i.e. $\eta=1$. Our simulations show an acousto-optic overlap as $\eta_\s{\s{waveguide}} \approx 1/4$.

By substituting all these contributions to Eq. \ref{eq:BrillouinGain}, we obtain a theoretical 342-times gain reduction in \sini\ waveguides compared to single-mode fibers. The Brillouin gain value measured in our \sini\ waveguide, in units m/W, is 250-times smaller than the silica intrinsic gain ($\tilde{g}_{\s{B, Si_3N_4}} = \tilde{g}_{\s{B, SiO_2}}/250$), in agreement with the theoretical value derived above. When the intrinsic gain is considered, however, phonon leakage is absent and acousto-optic overlap is unity ($\eta=1$). Therefore, the intrinsic \sini\ Brillouin gain is estimated as $7\times 10^{-13}~\s{m/W}$, 30-times smaller than that of silica.

In addition, the SBS threshold \cite{boyd1990noise} is estimated as 87 kW (see Supplementary Material). To give two comparisons, a 1-m-long standard single-mode fiber has a SBS threshold of 100 W and a hypothetical silica waveguide with the same dimensions as our waveguide would have a SBS threshold power of 20 kW.

\textit{Conclusion - }We have characterized backward SBS in integrated \sini\ waveguides. 
The observed SBS gain spectrum exhibits multiple peaks due to the hybridization with bulk acoustic resonance modes in the presence of SiO$_2$ cladding of finite thickness.
The calculated acoustic velocity in \sini\ from the measured Brillouin frequency shift (Eq. \ref{eq_phase_matching}) agrees with the reported value in Ref. \cite{Wolff:14}. Note that the observed SBS frequency shift in \sini\ of 25 GHz is the largest SBS frequency value reported on integrated platforms \cite{eggleton_brillouin_2019}. In addition, the SBS threshold in \sini\ is estimated as 87 kW in our 5-mm-long waveguide. Since SBS usually limits the maximum optical power in waveguides, its high threshold in \sini\ shows excellent high power handling capability of \sini\, central for integrated nonlinear photonics such as soliton microcomb \cite{kippenberg_dissipative_2018} and chip-based supercontinuum generation \cite{gaeta_photonic-chip-based_2019}.
Our work also allows assessing the fundamental noise associated with light propagation in Si$_3$N$_4$ waveguides as caused by thermal excitations of acoustic modes.

\noindent \textbf{Acknowledgments}: This work was supported by the Swiss National Science Foundation (SNSF) under grant agreement No. 159897, 178895, 176563 (BRIDGE), and by funding from the European Union’s H2020 research and innovation programme under FET Proactive grant agreement No. 732894 (HOT), and by the Defense Advanced Research Projects Agency (DARPA), Microsystems Technology Office (MTO) under contract No. HR0011-15-C-0055 (DODOS). J.H. acknowledges the support provided by Prof. Hwa-Yaw Tam and from the GRF of the Hong Kong Government under project PolyU 152207/15E. The Si$_3$N$_4$ samples were fabricated in the EPFL center of MicroNanoTechnology (CMi).

\noindent \textbf{Data Availability Statement}: The code and data used to produce the plots within this work will be released on the repository \texttt{Zenodo} upon publication of this preprint.

\bibliographystyle{apsrev4-2}
\bibliography{bibliography}

\end{document}


\maketitle

\tableofcontents

\section{Challenges, quantitative arguments}

In our waveguide, the silicon nitride (\sini) Brillouin gain has been measured to be $g_B =$ 0.07 m$^{-1}$W$^{-1}$. Thus the probe beam, just after having exited the waveguide, has a Brillouin-induced maximum intensity variation of the order of:
\begin{equation}
G_B\approx g_{B}P_{P0}L\sqrt{\alpha_{\s{wg}}} \approx 1.6\times 10^{-5},
\label{eq_gain_om}
\end{equation}
where $P_{P0}\approx 100~\s{mW}$ is the pump power inside the waveguide, $\alpha_{\s{wg}}\approx 0.2$ (7 dB) is the waveguide insertion loss and $L\approx 5~\s{mm}$ is the waveguide's length. As example, for a probe power of 0 dBm at detection, the stimulated Brillouin scattering (SBS) signal power at the gain peak is at -48 dBm. As a consequence, any noise at detection exceeding this value will inevitably cover the Brillouin signal. We now proceed to quantitatively describe the noise sources that we encountered in the experiments.

\subsection{Noise induced by polarization and temperature fluctuations}
\label{CWPumpProbe}
\begin{figure}[htbp]
\centering
	\includegraphics[width = 15cm]{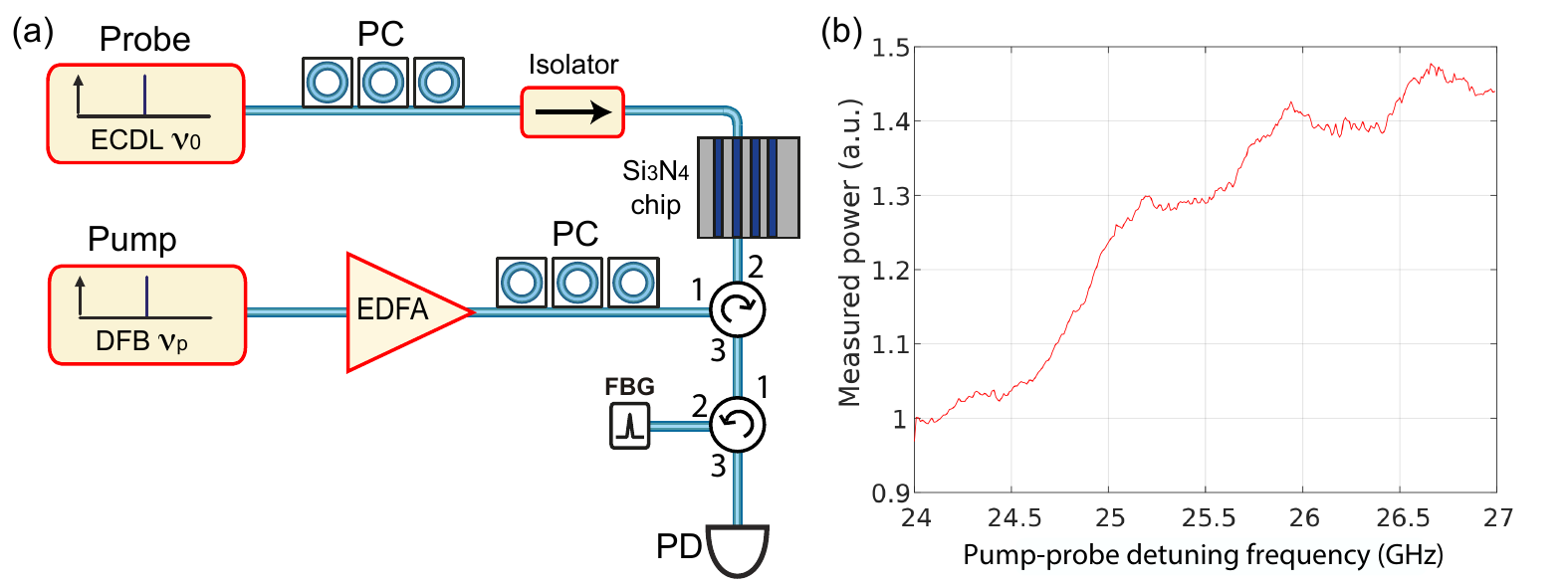}
	\caption{\textbf{CW pump-probe experiment.} \textbf{(a)} Experimental setup. \textbf{(b)} Measured signal showing a huge random noise. The tiny \sini\ Brillouin gain is fully covered by this environmental noise. ECDL, external-cavity diode laser; PC, polarization controller; EDFA, erbium-doped fiber amplifier; DFB laser, distributed feedback laser; FBG, fiber Bragg grating; PD, photodetector.}
	\label{fig_ch_CW}
\end{figure}

When a continuous wave (CW) pump-probe experiment, such as the one presented in Fig. \ref{fig_ch_CW}(a), is plugged to the \sini\ chip, the measured probe power exhibits a very noisy behaviour. A typical representation of this noise is given in the measurement of Fig. \ref{fig_ch_CW}(b). Each acquisition leads to a random trace profile similar to the one shown. This noise is generated by random thermal/polarization fluctuations modulating the fraction of pump power being coupled to the cavity, which in turn dictates the waveguide's temperature. These fluctuations modulate the probe transmission and the resulting huge probe variation compared to the small gain value to be measured prevents any CW experiment to be conclusive in measuring the Brillouin gain.

\subsection{Pump reflection}
Power reflection from the chip coupling points is experimentally measured to be $R_{\s{wg}} \approx -22$ dB. This means that 0.5\% of the pump power returns back and directly arrives to the detector in the absence of optical filtering. Apart from the danger of breaking the detector when high pump powers are used (e.g. 1 W), the noise induced by this reflection totally covers the Brillouin gain signal. Thus, the pump beam needs to be optically filtered out from the probe with the help of fiber Bragg gratings (FBGs). In the triple intensity modulation setup, we further filtered out the pump signal in the radio-frequency (RF) domain by modulating both pump and probe beams at a slightly different frequency and detecting the signal at the frequency difference, as explained in the main manuscript. This filtering method is very effective and totally removes the influence of the pumps from the measurement signal.

\subsection{Cavity Kerr effect}

\begin{figure}[t!]
\centering
	\includegraphics[width = 10cm]{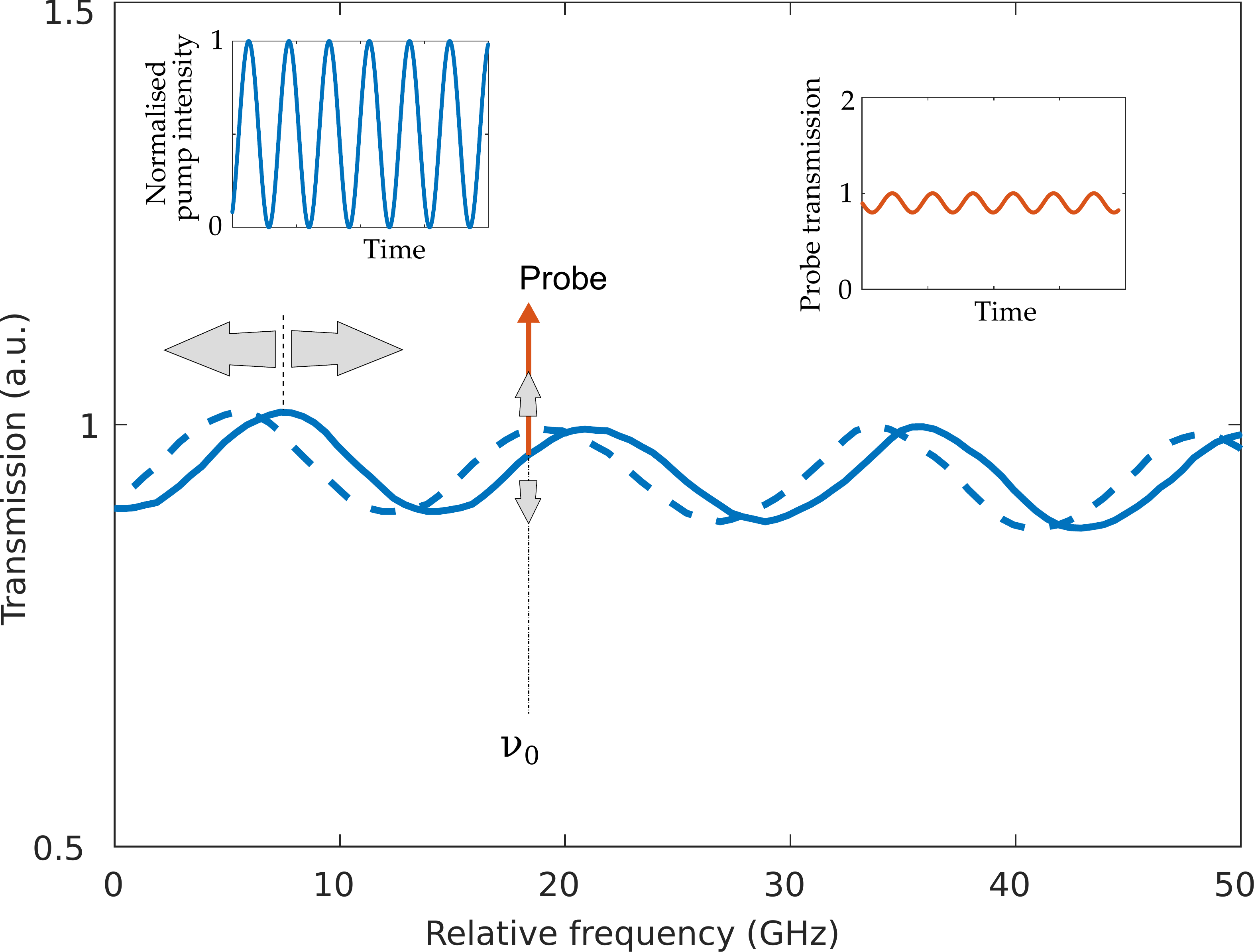}
	\caption{\textbf{Illustration of Kerr modulation.} An intensity-modulated pump periodically shifts the probe transmission spectrum, which in turns modulates the probe signal at the pump frequency. The solid blue trace corresponds to the measured waveguide transmission spectrum in function of a relative frequency starting at 193.5 THz. The dotted line represents this same spectrum shifted in frequency by pump-induced Kerr effect. Note that the shift is exaggerated for clarity.}
	\label{fig_ch_KerrModulation}
\end{figure}

Probe modulation by Kerr phase-to-intensity conversion induced by an intensity-modulated pump is depicted in Fig. \ref{fig_ch_KerrModulation}. We now compute the probe modulation amplitude in the worst case scenario (which occurs at random times due to random temperature/polarization fluctuations). The waveguide-induced cavity free spectral range (FSR) is:
\begin{equation}
\nu_F \approx \frac{c}{2n_gL},
\end{equation}
where $n_g = 2.1$ is the group refractive index at 1550 nm wavelength, $c$ is the speed of light in vacuum and $L$ is the waveguide length. The change in FSR occurring due to Kerr effect with respect to the pump power is:
\begin{equation}
\frac{\partial \nu_F}{\partial P} \approx -\frac{c}{2n_g^2L}\frac{\partial n_g}{\partial P},
\end{equation}
for which
\begin{equation}
\frac{\partial n_g}{\partial P}=\frac{\partial n}{\partial P}=\frac{n_2}{A_{\s{eff}}},
\end{equation}
where $n_2 = 2.5\times 10^{-19}$ m$^2$W$^{-1}$ is the \sini\ Kerr coefficient \cite{herr2013solitons} and $A_{\s{eff}} = 1.2~\s{\upmu m^2}$ is the waveguide's effective area. The FSR change is given by:
\begin{equation}
\Delta\nu_{F, \s{MAX}} = \frac{\partial \nu_F}{\partial P}\Delta P = -\frac{n_2c}{2n_g^2LA_{\s{eff}}}\Delta P\approx -283~\s{Hz},
\end{equation}
where, in our case, $\Delta P = 200$ mW. So the maximum cavity spectrum shift at the optical frequency $\nu_0$ is:
\begin{equation}
m\cdot |\Delta\nu_{F, \s{MAX}}|\approx 4~\s{MHz},
\label{result_cavity_shift}
\end{equation}
where $m = \nu_0/\nu_{F}\approx 13535$. Now, we calculate the transmission spectrum variation for the probe.
The waveguide's induced cavity transmission spectrum can be expressed as \cite{SalehTeich2007}:
\begin{equation}
P_N = \frac{1}{1+\left(2\frac{F}{\pi}\right)^2\sin^2{(\pi\frac{\nu}{\nu_F})}},
\end{equation}
where $P_N$ is the normalized transmitted power and $F$ is the cavity finesse. In our case, $\nu_F = 14.375$ GHz. Since the finesse is low, it can be simplified to:
\begin{equation}
I_N \approx 1-2\frac{F^2}{\pi^2}+2\frac{F^2}{\pi^2}\cos{\left(2\pi\frac{\nu}{\nu_F}\right)}.
\label{eq_finesse_simpl}
\end{equation}
Referring to the measured transmission spectrum shown in Fig. 2(a) of the main manuscript, the finesse can be found to be: $F \approx 0.53$. The maximum slope of (\ref{eq_finesse_simpl}) is:
\begin{equation}
S_{\s{MAX}} = \max_{\nu}\left\{\frac{\partial I_N}{\partial \nu}\right\} = \frac{4F^2}{\pi\nu_F}.
\end{equation}
In our case, we have $S_{\s{MAX}} \approx 24.9$ THz$^{-1}$. 
Combining this last result with (\ref{result_cavity_shift}), we obtain a worst-case probe variation of $G_K \approx 10^{-4}$, which is more than six times higher than the targeted Brillouin signal of Eq. (\ref{eq_gain_om}). In an idealized situation, this generated signal should be constant and only contribute to an offset. However, the cavity spectrum random shifts with temperature and evolution of pump coupling to the chip results in a random change of the offset and generates a noise of standard deviation two times larger than the peak Brillouin gain.

\subsection{Stimulated Raman scattering}
At a pump-probe detuning frequency equal to the Brillouin frequency shift, $\Delta\nu =\nu_B =$ 25 GHz, the silica Raman gain can be estimated to be \cite{stolen1989raman}:
\begin{equation}
\tilde{g}_{R}(\nu_B) \approx \tilde{g}_{R, \s{MAX}}\cdot 1.8\times10^{-3} \approx 10^{-16}~\s{m/W},
\end{equation}
where $\tilde{g}_{R, \s{MAX}} = 6.5\times 10^{-14}$ m/W is the peak silica Raman gain. Our patchcord length connecting to the waveguide is $L_{\s{pc}} =1$ m for each side. Given that the chip insertion loss is $\alpha_{\s{wg}}\approx 0.2$ (7 dB) and by taking into account the two coupling patchcords as well as the 1-cm-long high numerical aperture (NA) fibers used to couple the waveguide, the total effective length is: $L_{\s{pc, eff}}\approx$ 2.4 m.

Taking single-mode fiber effective area to be $A_{\s{eff,SMF}}\approx 80$ $\upmu$m$^2$ and considering that the pump power inside the patchcord is $P_{P1, \s{SMF}}$ = 24 dBm, the total stimulated Raman scattering occurring in the patchcord is:
\begin{equation}
G_R \approx \frac{\tilde{g}_{R}(\nu_B) }{A_{\s{eff,SMF}}}\cdot L_{\s{pc, eff}}\cdot  P_{P1, \s{SMF}} \approx 10^{-6}.
\end{equation}
Hence, in the present case, this background signal is smaller than the peak Brillouin signal and, despite contributing to the total noise, it does not cover the \sini\ peak Brillouin gain. Nonetheless, it is reduced thanks to the use of the triple intensity modulation experimental setup, as described in the main manuscript.

\section{Sample fabrication}
The integrated low-loss \sini\ waveguides are fully buried in SiO$_2$ cladding. The \sini\ waveguide sample presented in this work is fabricated using the subtractive process \cite{Luke:13, Xuan:16}. In this process, the \sini\ film from low-pressure chemical vapour deposition (LPCVD) is first deposited on the thermal wet oxide substrate on a 4-inch silicon wafer. Electron beam lithography (EBL) is used to pattern the waveguides, followed by dry etching using CHF$_3$/O$_2$ gases which transfers the waveguide pattern from the EBL resist mask (HSQ) to the SiO$_2$ substrate. Afterwards, the entire wafer is annealed at 1200$^\circ$C to drive out the residual hydrogen content \cite{Liu:18a} in \sini\, which can cause strong light absorption losses. Top SiO$_2$ cladding of 2.3 $\upmu$m thickness is then deposited on the substrate, followed by SiO$_2$ thermal annealing at 1200$^\circ$C once again. Due to the added height of the waveguide before SiO$_2$ cladding deposition, a dome-shaped top silica-air boundary is formed after deposition, as shown in the simulation results of Fig. 3 of the main manuscript.  Finally, the wafer is separated into chips of $5\times5$ mm$^2$ in size via dicing or deep dry etching.

Inverse nanotapers are used to couple light both into and out of the chip via high numerical aperture (NA) fibers \cite{Liu:18}. The coupling loss is less than 2 dB/facet, corresponding to a fiber-chip-fiber through coupling efficiency of $40\%$. The high NA fibers are packaged to the \sini\ chip \cite{Raja:19a}, which allow compact, portable devices for transfer and easy integration into a fiber system.

\section{Error bar estimation}
The uncertainty in the gain estimation comes from the imprecise knowledge of four key-parameters. As mentioned in the main manuscript, the measured peak Brillouin gain in units of $\s{m^{-1}W^{-1}}$, $g_\s{B}$, can be written as:
\begin{equation}
g_\s{B} \propto \frac{|A_{\s{det}}|}{P_\s{S} \cdot P_\s{P} \cdot \rho_\s{pd} \cdot L},
\end{equation}
where $|A_{\s{det}}|$ is the detected signal magnitude. Thus, the uncertainty of $g_\s{B}$ depends on the uncertainty of the following parameters:

\begin{table}[htb]
	\centering
	\begin{tabular}{llrll}
		Symbol&Description&Value&&Unit\\\hline
		$P_\s{S}$&Time-averaged probe power at detection&-12.15&$\pm0.15$&dBm\\
	    $P_\s{P}$&Time-averaged pump1 power inside the waveguide&20.8&$\pm$1&dBm\\
	    $\rho_{\s{pd}}$&Photodetector power-to-voltage conversion factor&6.14&$\pm$0.35&kV/W\\
		$L$&Waveguide effective length, including inverse nanotapers &5&$^{+0}_{-0.3}$&mm\\	
	\end{tabular}
    \label{table_parameters}
\end{table}
Explanation of the error bars: $P_\s{S}$ uncertainty comes from standard deviation of repeated measurements. $P_\s{P}$ uncertainty comes from the uncertainty about the balance of the coupling loss on each side of the waveguide, that is estimated to be $\pm 1$ dB. $\rho_{\s{pd}}$ uncertainty comes from standard deviation of repeated measurements. $L$ uncertainty comes from the lack of knowledge about the nanotapers contribution to the total Brillouin gain, that is estimated to be 5$^{+0}_{-0.3}$ mm.
Assuming Gaussian distributions and applying propagation of errors, the resulting uncertainty for the Brillouin gain, $g_\s{B}$, is $(0.07\pm0.01)$ m$^{-1}$W$^{-1}$ and $\tilde{g}_\s{B}$, is $(8\pm 1) \times 10^{-14}$ m/W. The uncertainty of the photoelastic constant $p_{12}$, is obtained by the fact that it is proportional to the square root of the Brillouin gain, giving a value of $|p_{12}|= 0.047\pm 0.004$. It is assumed that the additional error caused the simulation as well as the estimation of acoustic Q-factor (obtained from Lorentzian fitting of the Brillouin gain) and the effective area, is negligible.

\section{Simulations}
Simulations of the optical and acoustic modes are performed using COMSOL\textregistered\ 2D "Electromagnetic Waves, Frequency Model" and 3D "Solid Mechanics" modules, respectively. \sini\ refractive index $n_\s{Si_3N_4} = 2.00$ and SiO$_2$ refractive index $n_\s{SiO_2} = 1.45$ are used to compute the optical modes. \sini\ density $\rho_\s{Si_3N_4}$ = 3100 kg/m$^3$, \sini\ Young's modulus $E_\s{Si_3N_4} =$ 280 GPa \cite{pierson1999handbook}, SiO$_2$ density $\rho_\s{SiO_2} =$ 2203 kg/m$^3$ and SiO$_2$ Young's modulus $E_\s{SiO_2} =$ 73 GPa are used to compute the acoustic eigenmodes. These acoustic eigenmodes are computed by solving the equation of motion in frequency domain, including the strain-displacement relation and Hooke's law.  The \sini\ photoelastic constant $p_{12}$ is adjusted to match the Brillouin gain obtained from our simulations to the measurement data (see discussion). For each acoustic eigenmode, the Brillouin gain is obtained by computing the overlap integral between the eigenmode displacement field and the body forces resulting from the divergence of the electrostrictive (stress) tensor \cite{qiu2013stimulated}. Acoustic phonon leakage and subsequent reflections at the chip upper part (SiO$_2$) and at the silicon substrate are taken into account by including the SiO$_2$ cladding, the silicon substrate and a surrounding 1.5 $\upmu$m thick PML in the simulation (except for the upper part for which a free boundary is used to imitate the silica-air boundary). The acoustic damping is calculated from the measured Brillouin linewidth of 390 MHz. The simulation results are slightly shifted in frequency (220 MHz) to match the measurement results. This shift is likely the result of a slight mismatch of parameters in the simulation, such as Young's modulus or density.

\section{Threshold calculation}
To estimate the corresponding SBS threshold power in our \sini\ waveguide, only one pump of power $P_\s{P}$ and frequency $\nu_\s{P}$ is considered in the waveguide. The power reflection coefficient $R$, due to Brillouin back-scattering is given by \cite{boyd1990noise}:
\begin{equation}
R = Y \ex{G/2}\left(I_0\left(G/2\right) - I_1\left(G/2\right)\right),
\end{equation}
where $I_m$ are the modified Bessel functions of the first kind of $m$-th order, $G = g_\s{B}\cdot P_\s{P} \cdot L$ is the unitless gain with $L$ being the waveguide length. The parameter $Y$ is defined as:
\begin{equation}
Y = \frac{1}{4}(\bar{n}+1)g_\s{B}\cdot h\nu_\s{P}\cdot \Gamma \cdot L,
\end{equation}
where $h$ is the Planck constant, $\Gamma$ is the \sini\ acoustic damping rate and $\bar{n} = \left(\ex{h\nu_B/kT}-1\right)^{-1}$ is the mean number of phonons per acoustic mode (of frequency $\nu_B$) at temperature $T$, $k$ is the Boltzmann constant. The SBS threshold is conventionally defined as the required gain for $R(G_{\s{crit}})=0.1$. Therefore, the critical gain is estimated as $G_{\s{crit}} \approx$ 29 at room temperature, and the SBS threshold is estimated as $P_{\s{crit}}\approx 87$ kW.

\bibliographystyle{naturemag}
\bibliography{biblio_suppl}